\documentclass[runningheads,a4paper]{llncs}

\usepackage{epsfig,url}
\usepackage{amssymb}
\usepackage{amsmath}
\usepackage{amsfonts}
\usepackage{graphicx}
\usepackage{graphics}
\usepackage{color}
\usepackage{listing}
\usepackage{float}
\usepackage[ruled]{algorithm}
\usepackage{algpseudocode}
\usepackage{amssymb}
\usepackage{epsfig}
\usepackage{color}
\usepackage{xspace}
\usepackage{multirow}
\usepackage[multiple]{footmisc}

\usepackage[scaled=0.85]{couriers}

\usepackage[pdfpagelabels,
           plainpages=false,
           bookmarks=true,
           hypertexnames=false,
           pdffitwindow=false,
           colorlinks=true,
           linkcolor=blue,
           citecolor=blue,
           filecolor=black,
           urlcolor=black,
           pdfauthor={},
           pdftitle={},
           pdfsubject={2015},
           pdfkeywords={YYY},
           pdfproducer={Latex with hyperref},
           pdfcreator={pdflatex}
          ]{hyperref}

\usepackage[table]{xcolor}
\long\def\comment#1{}

\definecolor{red}{rgb}{1,0,0}

\urldef{\mailsa}\path|{luca.vassio,hassan.metwalley,danilo.giordano}@polito.it|
\newcommand{\keywords}[1]{\par\addvspace\baselineskip
\noindent\keywordname\enspace\ignorespaces#1}

\title{The Exploitation of Web Navigation Data:\\Ethical Issues and Alternative Scenarios}	
\titlerunning{The Exploitation of Web Navigation Data}

\begin{document}

\author{Luca Vassio \and Hassan Metwalley \and Danilo Giordano}

\authorrunning{Vassio et al.}

\tocauthor{Luca Vassio, Hassan Metwalley, Danilo Giordano}

\institute{Dipartimento di Elettronica e Telecomunicazioni \\ Politecnico di Torino, Italy\\
\mailsa}

\maketitle

\begin{abstract}
Nowadays, the users' browsing activity on the Internet is not completely private due to many entities that collect and use such data, either for legitimate or illegal goals. 

The implications are serious, from a person who exposes unconsciously his private information to an unknown third party entity, to a company that is unable to control its information to the outside world. As a result, users have lost control over their private data in the Internet. 

In this paper, we present the entities involved in users' data collection and usage. Then, we highlight what are the ethical issues that arise for users, companies, scientists and governments. Finally, we present some alternative scenarios and suggestions for the entities to address such ethical issues. 


\end{abstract}

\section{Introduction}
\label{sec:currentScenario}

Nowadays, almost everybody use the World Wide Web for different reasons, such as looking for news and products, accessing social networks and organizing their lives. Also in a business environment, most of the companies can not run their business without exploiting the web.
For these reasons, the content that a user is consulting on the web can be classified as sensitive personal information, either if the user is a company or a private person. 
Thanks to the design of the web, many entities can access, partially or totally, these sensitive information. 
Internet Service Providers (ISPs), online services, social networks and trackers usually store as much information as possible about the users.

The data that are collected can be used in many ways: ISPs could use these information to improve their network and the quality of their services; scientists and researchers could use these data to design new services or for understanding web social implications; advertisement companies could profile users to give them specific ads; criminals could steal identities or bank information; police and public agencies could find proofs and incriminate people; companies could know what their employees are doing and decide whether to hire someone or to fire someone else.

It is clear from this partial list that the exploitation of users' web navigation data is a very complex and delicate topic, where laws are still not comprehensive and many entities are not even aware of the current scenario.

The reminder of this paper is structured as follow: in Section~\ref{sec:entities} we identify and deeply explain the role of all the stakeholders in the current scenario, highlighting the connections among them. In Section~\ref{sec:ethical_issues} we present the  ethical issues that arise for the different groups of entities. In Section~\ref{sec:alternative_scenarios} we suggest some alternative scenarios, suggesting to the different entities how they could behave and some possible counter-measures to avoid improper use of users' sensitive data. Finally, some concluding remarks are presented in Section~\ref{sec:conclusions}.
\section{The stakeholders network in the current scenario}
\label{sec:entities}

\begin{figure*}[t]
\begin{center}
 \includegraphics[width=0.95\textwidth]{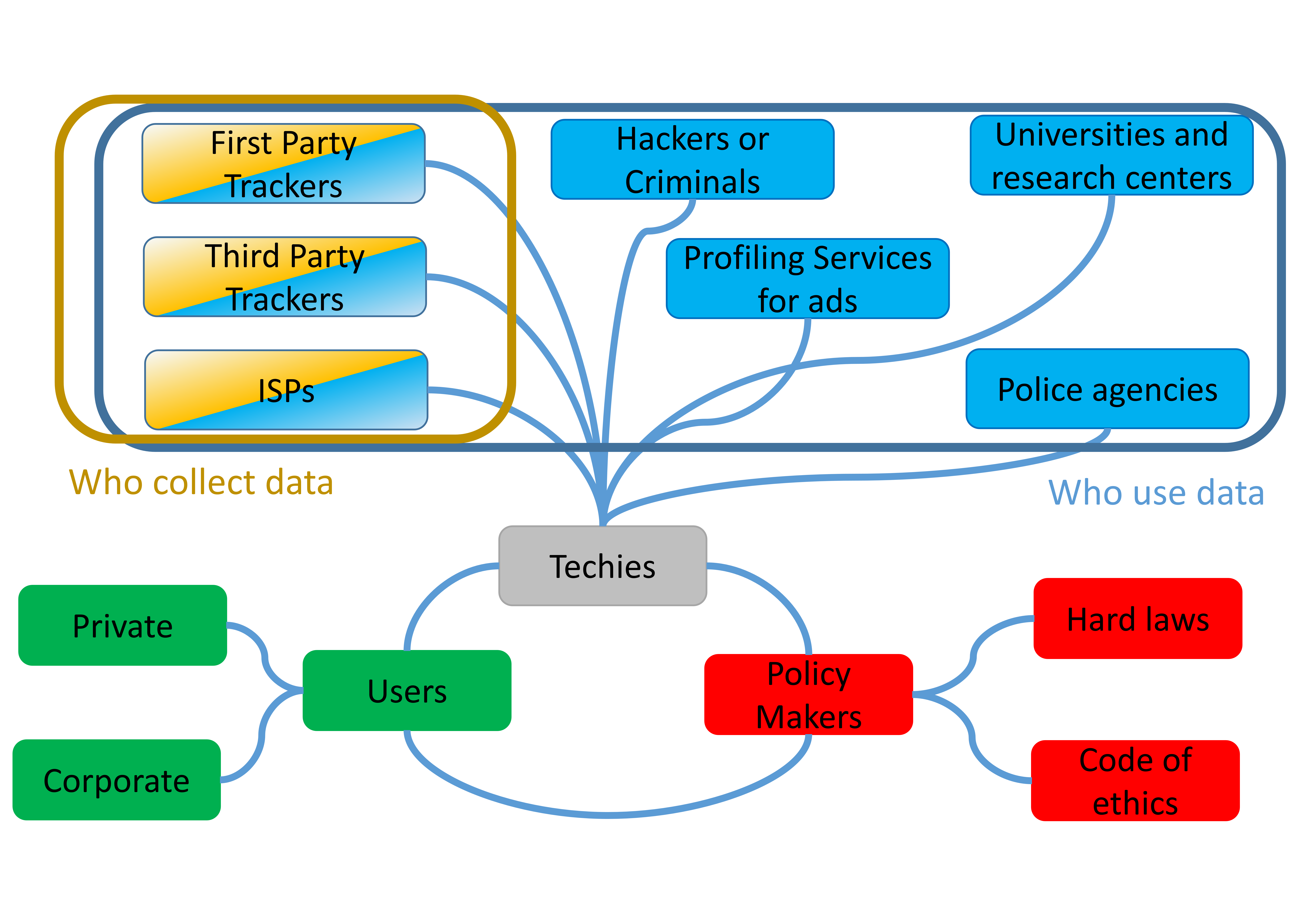}
 \caption{The stakeholders network involved in producing, collecting, using and controlling the web navigation data.} 
 \label{fig:entities}
\end{center}
\end{figure*}

The \textit{stakeholders  network} has been proposed~\cite{Stakeholders05} as a powerful tool for analyzing and reasoning about the difficult choices within an ICT scenario. 
The simple construction of this network is already a good help to identify conflicts between stakeholders and missing relationships usually not considered into the specific studied landscape. 
The \textit{stakeholders network} for the web navigation data scenario is presented in Figure~\ref{fig:entities}. We will explain all the key actors of the network, as well as the connections between them. 

The actors in the scenario are called \textit{entities} and can be grouped in different families, based on the role they play. 

The first group of entity is the \textit{users} family, depicted with green background in Figure~\ref{fig:entities}. The users are the part of the network that generates data by browsing on the web; therefore they are the entities that are potentially more exposed to risks. This dangerous position is often followed by an unawareness of such risks. Indeed, as we will explain later, only a small percentage of users use some techniques to protect their data. Users family can be further divided into two entities: \textit{private} and \textit{corporate}. Private users generate traffic for personal reasons, e.g., web surfing, entertainment or gaming. These users do not wish to disclose their data both for privacy and for security reasons; for example, they would not to share personal interests or banking information. Corporate users, instead, generate traffic for business reasons: security of these data should be fundamental for companies to avoid industrial espionage like market analysis and information about new projects.

The second group of entities is the one that could access the data, represented in yellow in Figure~\ref{fig:entities}. Within this family, each entity has a different view of the data and owns them for different reasons. 

The first entity is composed by the \textit{Internet Service Providers} (\textit{ISPs}). They give to the users the physical access to the network, therefore an ISP receive all traffic from its users. As a consequence, ISPs are the entities that can see everything about users behaviour, except for the encrypted traffic, i.e., HTTPS. 

Secondly, there are the so called \textit{First Party Trackers}. This entity is composed by service providers, such as: \textit{Google}, and \textit{Facebook}. They receive some users' data directly from the users in order to receive a particular service. For example, an internaut that contacts \textit{YouTube} to retrieve a video has to send the information about the video that wish to stream in order to watch it. 

The most active, and controversial entry of this group of entities is composed by the \textit{Third Party Trackers}. These services are embedded into the websites and are not directly linked to the original source being asked. They monetize visits and, in practice, collect users' personal information. Trackers use many solutions to identify users, ranging from storing cookies on the user's browser to tracking techniques that fingerprint users across several web sites~\cite{acar2014web,Bala2011,Rezgui2003}. Google’s DoubleClick, and Yahoo YieldManager are notable examples. However, the list of companies that build their business around information collection is of the order of several hundreds of entries. A recent work based on passive measures~\cite{HassnTMA15}, counted more than 400 active online tracking services. With 100 of them being regularly contacted by more than 50\% of users, and the most pervasive ones that are impossible to avoid. In addition, in this work the authors demonstrates that 77\% of users faced the first tracker just 1 second after starting their online web activity. This demonstrates that this phenomenon is enormously popular and it involves all Internet users.

The third group in the stakeholders network is composed by entities that wish to use users' data, depicted in blue in Figure~\ref{fig:entities}. Inside of this family there are either entities that could be legitimate to use some users' data, either others that wish to exploit them for malicious activities. As we can see from Figure~\ref{fig:entities}, this family includes all the three entities that can directly collect users' data (ISPs, First, and Third Party Trackers).

Nowadays, as explained before, ISPs have the best position to collect users' data. However, even if they have such a privileged position, they exploit them only for few operations. Currently, due to the birth of Software Defined Network (SDN), ISPs start offering new services which can heavily exploit private data. Before this scenario, ISPs already exploited users' data to extract some information to improve the users' Quality of Experience (QoE). An example is given by the so called \textit{transparent caching} or \textit{transparent proxy}~\cite{TransparentProxy}. ISPs try to understand what are the most popular content requested in their network in order to cache such contents inside some of their servers. By applying this technique, ISPs can provide smaller latency for such contents, therefore improving users' QoE. Clearly, this solution allows ISPs to save money as well. Indeed, storing contents inside of their servers is cheaper with respect to send the requests via other networks. In fact, ISPs have a non-marginal cost for sending data through networks not directly owned by them.

First Party Trackers use private information to provide some services. In addition to the needed informations, they also collect other data like cookies to allow users' login, or to remember the basket in case of e-commerce websites. However, they could also exploit such information to improve their services, targeting them to a specific subset of users. For instance, many websites target specific e-mails for advertisement of their products or services on the basis of the users' interest shown in their website. 

The third party trackers are the most ambiguous since they base their business directly on the collection of personal information. Information that can be  extracted either implicitly, either explicitly from the users' web browsing activity. These companies directly use the data and could also sell them. This phenomenon is ubiquitous, with all major players and hundreds of mostly unknown companies taking part in it. The research community has focused on disclosing and quantifying the vastness of this problem~\cite{acar2014web,Bala2011,Yen2012,HassnTMA15}, but proposing just few solutions~\cite{Agarwal:2013:PDM:2462456.2464460,Enck:2010:TIT:1924943.1924971}.

As a matter of facts, profiling services are fundamental for marketing purposes, e.g., knowing products browsed on a shopping website, online newspapers usually read or movies liked. These information are used to deliver customized ads for the specific user. First and third party tracker themselves, other than companies that specifically buy these information, usually exploit the data for this purpose. 

Another important usage of such data are police agencies and authorities: as a matter of fact in many cases they access these data for mass surveillance. Despite this practice can be positive, as it could improve the safety of people, in many cases the authorities have overstated, taking too much information for the original purposes~\footnote{The {NSA} Leaks Are About Democracy, Not Just Privacy,\\ \url{http://www.theatlantic.com/politics/archive/2014/01/the-nsa-leaks-are-about-democracy-not-just-privacy/282902/}}$^{,}$\footnote{Man behind {NSA} leaks says he did it to safeguard privacy, liberty,\\ \url{http://edition.cnn.com/2013/06/10/politics/edward-snowden-profile}}. An important public example of such trend is reported by Google with the Google \textit{Transparency Report}.\footnote{\url{https://www.google.com/transparencyreport/}} Indeed, by visiting this report page, it is possible to discover how many information have been requested by government agencies and how many of them have been actually disclosed by Google.

Also researchers in companies and universities are interested in using such data for improving the knowledge in different topics by applying data-mining and statistical techniques~\cite{kosala2000web}. For example, researchers studies can be useful to find the social implications of the Web or for designing better technologies to surf the web.

Last but not least, since these data are stored somewhere in data clusters, criminals could break these systems and steal huge amount of personal data. These information can illegally be used to steal identities, bank information~\footnote{Bank Hackers Steal Millions via Malware, \url{http://www.nytimes.com/2015/02/15/world/bank-hackers-steal-millions-via-malware.html?_r=0}}, or passwords~\footnote{LinkedIn investigating reports that 6.46 million hashed passwords have leaked online, \url{http://www.theverge.com/2012/6/6/3067523/linkedin-password-leak-online}}. A historical important case of malicious activity is related to a doctor who sold a list of almost 4,000 HIV patients due to a grudge against his company~\footnote{Man convicted of breaking patient confidentiality, \url{http://www.ahcmedia.com/articles/38091-man-convicted-of-breaking-patient-confidentiality}}. 

The last group of entities in the stakeholders networks consists in the policy makers, depicted in red in Figure~\ref{fig:entities}. Hard laws are mandatory policies with legal values. In this family fall UE and US government laws, as well as regional ones. Soft laws are instead the ones that are not enforced and are related to ethical issues: computer engineers, as well as general users and companies, have their own code of ethics and privacy policies.
\section{Ethical and social implications: open questions}\label{sec:ethical_issues}

As described by Walter Maner about twenty years ago~\cite{Maner95}, the use of computing technology creates, and will create, novel ethical issues that require specific studies. 
In this section will be presented some ethical issues and social implications that arise for the different groups of entities introduced in Section~\ref{sec:entities}. We will arise some open questions, that do not have strict and definitive answers. Later on, in Section~\ref{sec:alternative_scenarios} we will give some advices to address these questions.

The ethical issues implicated in the users' web data usage belong the following two domains:

\begin{enumerate}
\item \textbf{Privacy:} data related to navigation history are certainly confidential and sensitive. Therefore, is it possible to use and store users' navigation data without violating their privacy? How the different entities should behave in order to respect this fundamental right?
\item \textbf{Computer crimes:} hackers can use their skills to access the stored data for malicious activities. Thus, how can entities protect these data and be sure not to put in danger users? For how long entities should store users' data?
\end{enumerate}

As we saw before, each entity has different reasons to save and use users' data. Despite of the reasons, the first questions are related on the way they obtained the data: did they receive an explicit or an implicit  user's consent, to store and save data? Was the user aware of being tracked? Many times the answer to both question is simple: the user was not conscious of the situation, therefore he did not give any consent at all. This is often true even in presence of an explicitly form that aware the user of the data collection. For instance, many websites present policy terms to the users that have to be explicitly accepted. However, often users do not even read these forms, since they look standards, they are long, and not easily understandable. Moreover, most of the time the user has the perception to be constrained in order to receive the service he is interested in. As a result, how can an entity be ethically legitimate to record users' data?

Starting from these points several other questions arise for the different entities.
\begin{itemize}
\item Should entities collecting users' data share them with other entities? In the case that the second entity is a state police agency that wish to use such data to prove a crime, is the answer the same? Even in the case where such police agency is not of the same country of the entity? In an extreme case, if in the police agency country there is the death penalty for the crime committed, what will be ethical? What should the involved entities do?
\item Many companies use users' data to profile users for advertisement goals. How deep can they go into the profiling? Is there a limit due to privacy? In this case, where is this limit and who choose it? For example, is it legitimate to send advertisements related to the health-care of a person?
\item Universities and research centres desire to use as much data as they can, to be able to perform better studies. However, is it fair that they could see personal data? Who can access the database where the data are stored? For what exact purpose? For example, is it an acceptable goal to study specific users' interests? 
\item A global network such as the internet makes usual physical boundaries obsolete. How can regional hard laws face trackers that are outside their jurisdiction? How can the freedom and anarchy of the web be balanced with a centralized regional control?
\end{itemize}

These are only a limited part of the possible questions that arise by analyzing the current internet scenario, for which fair definitive answers do not exist. However, we hope that by just proposing these questions, we can stimulate a debate between the parties that are involved in today internet network.
\section{Present and future possible alternative scenarios}\label{sec:alternative_scenarios}

In Section~\ref{sec:ethical_issues} we identified the ethical issues related to users' web navigation data. We will now see what kind of advices and suggestions we can give to the entities of the stakeholders network presented in Figure~\ref{fig:entities}.

\subsection{Users}

An internet user cannot avoid to use an ISP to access the internet. However, not all the ISPs are equal. A user has the possibility, at least, to take a look at the specific ISP policy before starting a new contract. Moreover new ISPs that want to make customer privacy their top priority are emerging~\footnote{{N}ew {ISP} {T}o {M}ake {C}ustomer {P}rivacy {I}ts {T}op {P}riority, \url{http://www.themarysue.com/privacy-first-isp}}.

Similarly, a user cannot simply block first-party trackers. However, if a user wants to still have a specific kind of service, he has the choice of moving to alternative service providers.
For search engines, a possibility would be to use search engines that do not track and log any personally identifiable information. For example, DuckDuckGo\footnote{\url{https://duckduckgo.com/}} does not use cookies to identify users, and it discards user agents and IP addresses from its logs. Moreover it does not even generate anonymized identifier to tie searches together. Therefore, the search engine has no way of knowing whether two searches even came from the same computer and you will get the same results as everyone else in the world. If a user still prefer Google’s search results, it can be possible to use services such as Startpage.\footnote{\url{https://startpage.com/}} This service submits your search to Google and returns the results to you. In this case, Google sees a large amount of searches coming from Startpage servers, without knowing who originally requests each content. Whether these approaches ensure your privacy, you will never have personalized search based on your interests. 

The are many possible counter-measures to avoid third party trackers to collect your data.
The privacy-conscious users have easily available many privacy-enhancer browser plugins. These plugins are actually the most used protection from the tracking services. They automatically block the transmission of user's identifying information, depending on the user's willingness. 
This type of software is directly installable into web browsers and permits user to easily modify traffic, i.e. disabling cookie sending or part of javascript page, or drastically block communications to certain web services. 
However, these solutions is ineffective for traffic generated out of browsers (e.g., mobile application) and, additionally, the diffusion of these solutions are surprisingly very limited, as shown by Metwalley~et~al.~\cite{HassnTMA15}. 
Despite end-users' concerns about privacy largely increased, motivated also by exposed government surveillance programs, Internet user does not fully grasp the extent and seriousness of the problem. To this end, a common misconception is that encryption of the web would help protecting users' privacy. Accordingly, HTTPS usage increased by 100\% each year, reaching about 50\% of web flows in October~2014~\cite{finamore:conext14}. In reality, encryption increases the value of data for third party trackers. Web services that deploy encryption establish a monopoly on information by precluding any other parties from exploiting it. Moreover, HTTPS prevents third parties and malicious users to check and possibly control what kind of data is exchanged. 
Another possible solution for the user would be the so-called \textit{Do Not Track HTTP header}.\footnote{\url{http://donottrack.us/}} It is an encouraging initiative that allows users to opt out of tracking by advertising networks and analytics services. With this solution the user can choose to turn on the field in his browser, that automatically sends a special signal to the web sites telling that the user would not like to be tracked. The main problem of this solution is that, currently, there is no consensus on how the companies you encounter should interpret the \textit{Do Not Track header}. As a result, most sites do not currently change their behaviour, with few sites supporting this solution~\footnote{\url{http://allaboutdnt.com/}}. 

\subsection{Hard laws}

The governments started recently to address the privacy and crime implications of the web navigation data. In 2011 the European Union~\cite{EU2011} stated that the web navigation data shall be obtained and processed fairly. This principle generally requires that a person whose data are processed to be aware of at least the following information:
\begin{itemize}  
	\item the identity of the person who is processing the data;
  	\item the purpose or purposes for which the data are processed;
 	\item any third party to whom the data may be disclosed;
  	\item the existence of a right of access and a right of rectification.
\end{itemize}
    
Another interesting initiative of European Union is the \textit{Cookie Law}~\footnote{The Cookie Law Explained, \url{http://www.cookielaw.org/the-cookie-law/}}. Thanks to this legislation, the websites must request consent from visitors to store or retrieve any information on a computer, smartphone or tablet. From June 2015 any website available to European visitors that uses cookies or any other technologies for non-essential tracking must:
\begin{enumerate}
	\item inform users that tracking technologies are used;
	\item explain the reasons for using those technologies;
	\item obtain the user’s consent prior to using that technology and allow them to withdraw permission at any time.
\end{enumerate}
While cookies are an obvious target, the law applies to all client-side technologies used to identify an individual. Additionally, user's consent must involve communication where the individual consciously indicates their acceptance, e.g., by clicking an icon or check box. The only exceptions are sites where tracking is strictly necessary for the provision of a service or communication requested by the user. Shopping baskets, some online applications and client-side caching to improve page speed would not require authorization. Instead, sites using analytics, advertising or customized greetings must comply.

In this context, an important problem is the difference, in terms of laws, between the European Union and the United States~\footnote{Differences between the privacy laws in the EU and the US, \url{http://resources.infosecinstitute.com/differences-privacy-laws-in-eu-and-us/}}. As a matter of fact, while the EU is trying to realize a set of laws dedicated to privacy, in the US the privacy regulation is based on a self-regulatory approach, where companies provide privacy notices that make certain promises about privacy. If these promises are violated, the \textit{Federal Trade Commission} (FTC) might penalize the company. But this power generally extends only to the promises made, so a company can determine how stringently it wants to protect privacy by modulating the promises it makes. In many instances, people are given only a right to opt out of certain uses of their data, and often no right at all to limit the collection of data about themselves by certain companies. In the EU, the rules regarding individual consent for data collection, use, and disclosures are much stricter, and much more affirmative consent is required.
Fortunately, United States government has started to collaborate with EU with the goal to improve users' privacy, and the first result of this cooperation are the \textit{International Safe Harbor Privacy Principles}~\footnote{Safe Harbor Frameworks, \url{http://www.export.gov/safeharbor/}}. These principles forbid to transfer personal data to non-European Union countries that do not meet the European Union Directive on the protection of personal data. The aims of these principles are the protection of personal data from accidental information disclosure or loss. This task represents the first step of a process that could lead to a common lawmaking with the aim of stopping, from the point of view of hard laws, the private information leakage.
   
\subsection{Researchers and Computer Scientists}

Computer professionals should follow an ethical code while dealing with personal information. The most famous ones are the IEEE and ACM code of ethics~\footnote{{IEEE code of ethics}, \url{http://www.ieee.org/about/corporate/governance/p7-8.html}}$^{,}$\footnote{{ACM} code of ethics, \url{http://www.acm.org/about/code-of-ethics} and \url{www.acm.org/about/se-code}}. In addition to these general-purpose code of ethics, recently the Oxford University  propose a set of guidelines for measurement projects regarding privacy~\cite{EthicGuideline}. However, the problem in this scenario is not simple since researchers and engineers must face two challenges: anonymize the data used during the research and study the services that steal private information. 

Regarding the first problem, some researchers~\cite{Chaudhuri2006,fan_prefix-preserving_2004} suggested different solutions to anonymize data, while maintaining only the information that could be useful for research purpose. However, it can be very difficult to predict if in the future records in supposedly anonymized datasets will be re-identified. In general, a scientist should modify and present data that are not privacy invasive. It may be possible to find a compromise in which some level of aggregation and pre-processing to de-identify the data takes place before a dataset is collected and stored.

On the other side, research community is currently searching a way to solve privacy problems together with companies and institutions, maintaining the economic ecosystem created around usage of users' information. Recently Telefonica, one of the most important private telecommunications companies in the world, have started an initiative called \textit{Data Transparency Lab}~\footnote{Data Transparency Lab, \url{http://datatransparencylab.org/}}. This is a collaborative effort between universities, businesses and institutions to support research in tools, data, and methodologies for shedding light on the use of personal data by online services, and to empower users to be in control of their personal data. This initiative represents a first example of how the research community can really solve privacy problems in this field.
\section{Conclusions}\label{sec:conclusions}

In this paper we have discussed, from a high-level point of view, ethical issues and social implications related to the users' web navigation data. We showed that in the current scenario many entities collect, store, and use these information, affecting the users' privacy. We presented that some entities might be legitimate to access these data, e.g., researchers or police agency, while other should not have any access at all, e.g., malicious hackers. 
However, since people will give even more importance to the web in the near future and the web will be more pervasive in everyday life, the ethical implications could even enlarge their effects. We have shown some alternative scenarios and some countermeasures to mitigate the problems that arise. Firstly, users should be really aware of how their data are used and how they could improve their privacy. Secondly, policy makers are slowly trying to regulate these phenomena, but they should be capable of fast interventions. Finally, even scientists and engineers should put their force to make these data harmless for the users, carefully evaluating the implications of the data usage for their projects.
\bibliographystyle{splncs}
\bibliography{computer_ethics}

\begin{thebibliography}{10}

\bibitem{Stakeholders05}
Gotterbarn, D., Rogerson, S.:
\newblock {Responsible Risk Analysis for Software Development: Creating the
  Software Development Impact Statement}.
\newblock {Communications of the Association for Information Systems} (2005)

\bibitem{acar2014web}
Acar, G., Eubank, C., Englehardt, S., Juarez, M., Narayanan, A., Diaz, C.:
\newblock {The Web Never Forgets: Persistent Tracking Mechanisms in the Wild}.
\newblock In: ACM SIGSAC. (2014)

\bibitem{Bala2011}
Krishnamurthy, B., Naryshkin, K., Wills, C.E.:
\newblock {Privacy leakage vs. Protection measures: the growing disconnect}.
\newblock In: W2SP. (2011)

\bibitem{Rezgui2003}
Rezgui, A., Bouguettaya, A., Eltoweissy, M.:
\newblock Privacy on the web: Facts, challenges, and solutions.
\newblock {IEEE} Security \& Privacy (2003)

\bibitem{HassnTMA15}
Metwalley, H., Traverso, S., Mellia, M., Miskovic, S., Baldi, M.:
\newblock The online tracking horde: a view from passive measurements.
\newblock In: TMA. (2015)

\bibitem{TransparentProxy}
Blum, S., Lueker, J.:
\newblock Transparent proxy server (2001) US Patent 6,182,141.

\bibitem{Yen2012}
Yen, T.F., Xie, Y., Yu, F., Yu, R.P., Abadi, M.:
\newblock {Host Fingerprinting and Tracking on the Web: Privacy and Security
  Implications}.
\newblock In: NDSS. (2012)

\bibitem{Agarwal:2013:PDM:2462456.2464460}
Agarwal, Y., Hall, M.:
\newblock {P}rotect{M}y{P}rivacy: {D}etecting and {M}itigating {P}rivacy
  {L}eaks on i{OS} {D}evices {U}sing {C}rowdsourcing.
\newblock In: {ACM} Mobi{S}ys. (2013)

\bibitem{Enck:2010:TIT:1924943.1924971}
Enck, W., Gilbert, P., Chun, B.G., Cox, L.P., Jung, J., McDaniel, P., Sheth,
  A.N.:
\newblock Taint{D}roid: {A}n {I}nformation-flow {T}racking {S}ystem for
  {R}ealtime {P}rivacy {M}onitoring on {S}martphones.
\newblock In: {USENIX} {OSDI}. (2010)

\bibitem{kosala2000web}
Kosala, R., Blockeel, H.:
\newblock Web mining research: A survey.
\newblock ACM Sigkdd Explorations Newsletter (2000)

\bibitem{Maner95}
Maner, W.:
\newblock {U}nique ethical problems in information technology.
\newblock In: ETHICOMP95, Leicester,UK (1995)

\bibitem{finamore:conext14}
Naylor, D., Finamore, A., Leontiadis, I., Grunenberger, Y., Mellia, M.,
  Papagiannaki, K., Steenkiste, P.:
\newblock The {C}ost of the {\textquotedblleft}{S}{\textquotedblright} in
  {HTTPS}.
\newblock In: {ACM} {C}o{NEXT}. (2014)

\bibitem{EU2011}

\newblock European Communities Regulations 2011, Electronic Communications
  Network and Services, Privacy and Electronic Communications. Statutory
  Instruments. S.I. No. 336 of 2011,
  \url{https://www.dataprotection.ie/documents/legal/SI336of2011.pdf}

\bibitem{EthicGuideline}
Zevenbergen, B.:
\newblock Ethical privacy guidelines for mobile connectivity measurements.
\newblock Oxford Internet Institute, University of Oxford,
  \url{http://www.themarysue.com/privacy-first-isp}

\bibitem{Chaudhuri2006}
Chaudhuri, K., Mishra, N.:
\newblock When random sampling preserves privacy.
\newblock In Dwork, C., ed.: Advances in Cryptology - CRYPTO 2006. Volume 4117
  of Lecture Notes in Computer Science.
\newblock Springer Berlin Heidelberg (2006)

\bibitem{fan_prefix-preserving_2004}
Fan, J., Xu, J., Ammar, M.H., Moon, S.B.:
\newblock Prefix-{Preserving} {IP} {Address} {Anonymization}:
  {Measurement}-{Based} {Security} {Evaluation} and a {New}
  {Cryptography}-{Based} {Scheme}.
\newblock Computer Networks (2004)

\end{thebibliography}

\end{document}